\documentclass[letterpaper,twocolumn,10pt]{article}
\usepackage{usenix-2020-09}

\usepackage{hyperref}

\usepackage{amssymb}
\usepackage{amsmath}
\usepackage{algorithm}
\usepackage{algorithmic}
\usepackage[capitalise]{cleveref}
\crefname{figure}{Fig.}{Figs.}
\Crefname{figure}{Fig.}{Figs.}
\crefname{equation}{Eq.}{Eqs.}
\Crefname{equation}{Eq.}{Eqs.}

\usepackage{lscape}
\usepackage{longtable}
\usepackage{tikz}

\usepackage{url}
\usepackage{booktabs}
\usepackage{multirow}
\usepackage{rotating}
\usepackage{xspace} 
\usepackage{array} 
\usepackage{subcaption}

\newcommand{\attackname}{APATE\xspace}%
\newcommand{\detectname}{MARRS\xspace}%

\begin{document}

\title{\Large \bf{Rogue Cell: Adversarial Attack and Defense in Untrusted O-RAN Setup Exploiting the Traffic Steering xApp}}

\author{
{\rm Eran Aizikovich}, {\rm Dudu Mimran}, {\rm Edita Grolman}, {\rm Yuval Elovici}, {\rm  Asaf Shabtai} \\
Ben Gurion University of the Negev
}
 \maketitle

\begin{abstract}

The Open Radio Access Network (O-RAN) architecture is revolutionizing cellular networks with its open, multi-vendor design and AI-driven management, aiming to enhance flexibility and reduce costs. 
Although it has many advantages, O-RAN is not threat-free.
While previous studies have mainly examined vulnerabilities arising from O-RAN's intelligent components, this paper is the first to focus on the security challenges and vulnerabilities introduced by transitioning from single-operator to multi-operator RAN architectures. 
This shift increases the risk of untrusted third-party operators managing different parts of the network.
To explore these vulnerabilities and their potential mitigation, we developed an open-access testbed environment that integrates a wireless network simulator with the official O-RAN Software Community (OSC) RAN intelligent component (RIC) cluster.  
This environment enables realistic, live data collection and serves as a platform for demonstrating \attackname (adversarial perturbation against traffic efficiency), an evasion attack in which a malicious cell manipulates its reported key performance indicators (KPIs) and deceives the O-RAN traffic steering to gain unfair allocations of user equipment (UE).
To ensure that O-RAN's legitimate activity continues, we introduce \detectname (monitoring adversarial RAN reports), a detection framework based on a long-short term memory (LSTM) autoencoder (AE) that learns contextual features across the network to monitor malicious telemetry (also demonstrated in our testbed).
Our evaluation showed that by executing \attackname, an attacker can
obtain a 248.5\% greater UE allocation than it was supposed to in a benign scenario.
In addition, the \detectname detection method was also shown to successfully classify malicious cell activity, achieving accuracy of 99.2\% and an F1 score of 0.978.

\end{abstract}

\section{Introduction}

In recent years, network operators and vendors have begun exploring innovative radio access network (RAN) architectures~\cite{demestichas20135g, bonati2020open}. 
The RAN provides wireless connectivity to mobile devices and acts as the final link between the cellular network and user equipment (UE).
Traditionally, RANs have been vendor-specific, i.e., with hardware interfaces and applications optimized for a specific vendor's equipment, and were operated by a single owner.
While this approach enables vendors to deliver integrated and highly optimized solutions, it also has significant drawbacks.
Traditional RANs require vendors to develop all components, driving up costs for network operators and creating vendor lock-in, which limits flexibility and innovation.
To overcome these drawbacks, an advanced architecture was proposed, the Open Radio Access Network (O-RAN)~\cite{polese2023understanding}.

O-RAN, which was introduced by the O-RAN Alliance\footnote{The O-RAN Alliance: \url{https://www.o-ran.org/who-we-are}}, started to gain attention, as it allowed:
(1) open accessible shared information to promote multi-vendor deployments, with standardized interfaces between RAN components~\cite{marinova2024intelligent,hasabelnaby2024centralized};
(2) adaptivity in real time by using cloud-based and virtualized components managed through software-defined networking (SDN), which enables more flexibility and reduces operational costs~\cite{vodafone2024,eugina2024ric}; and 
(3) adaption of the RAN intelligent component (RIC), which is responsible for utilizing artificial intelligence (AI) and machine learning (ML) systems, to reduce human intervention~\cite{niknam2022intelligent,parvez2018survey}.

While previous studies have begun to uncover vulnerabilities in O-RAN's intelligent components—particularly those exploited by malicious UE for personal gain~\cite{mimran2022evaluating}—the shift to multi-operators in RAN architectures introduces new security challenges.
Traditional RANs, where a single operator oversees the entire network, were considered fully trusted. 
However, the disaggregation of the O-RAN architecture has accelerated the shift toward multi-operator networks, resulting in an untrusted environment since different operators manage various network elements~\cite{otto2021oran,kee2021oran}.

In this paper, we demonstrate a threat model in which the threat actor is the cell (i.e., a cellular operator).
An adversarial cell can disrupt the network in several ways, such as reducing UEs migration to other cells or executing other denial-of-service (DoS) attacks.
These types of attacks can be financially driven, since operators may be compensated based on the amount of UE their cells serve.
To illustrate this threat, we introduce the \attackname (adversarial perturbation against traffic efficiency) attack.

The \attackname is an evasion attack~\cite{biggio2013evasion} targeting the traffic steering (TS) flow (part of O-RAN RIC use cases~\cite{WG3}) which is responsible for dynamically and intelligently managing network traffic. 
The attacker misleads the TS flow into assigning additional UE to its cell by falsely manipulating the TS's ML model that is responsible for the quality of experience (QoE) predictions (i.e., attacking the QoE predictor referred to as the QP).
The \attackname attack works as follows:
The attacker trains a substitute model of the QP model and uses this model to craft adversarial samples, i.e., samples that mislead the QP and cause it to make an incorrect prediction. 
The crafted adversarial samples are then used to query the QP target model, maliciously leading it to forecast an artificially high QoE for the attacker’s cell, by misleading the TS, resulting in an unfair allocation of UE to the malicious cell.

To mitigate these type of risks and ensure that O-RAN's legitimate activity continues, we propose a mitigation strategy called \detectname (monitoring adversarial RAN reports), specifically designed to detect untrusted actors attempting to disrupt legitimate network operations, such as in attacks like \attackname.
\detectname is a detection method based on a long short-term memory (LSTM) autoencoder (AE) architecture~\cite{lee2024lstm}. 
\detectname extracts relevant time-series features from the reporting cells and UEs key performance indicators (KPIs) and trains a dedicated AE model for each cell. 
Then, it uses the compressed latent space from each AE model, concatenated with the aggregated latent spaces from all other cells, to generate new enriched feature vectors that capture contextual information from both the specific cell and the entire network.
Next, an additional AE for each cell using the new enriched feature vector is trained, however this time in an attempt to reconstruct the original features.
Finally, a classifier is trained to compare the second model's output to the first model's input; if the loss between them exceeds a predefined threshold, then the input is classified as untrusted; otherwise it is classified as trusted.
The entire training process is performed using benign data to learn benign data behavior, i.e., unsupervised learning. 
At inference time, \detectname's classification will indicate whether the examined cell's KPIs are benign or adversarial.

To demonstrate the \attackname attack and evaluate the \detectname detection method, we developed a testbed with a wireless network simulator~\cite{de2022satellite} to emulate a live network topology with moving UE and gNBs (5G cell base stations) and an O-RAN Software Community (OSC) RIC cluster~\cite{bimo2022osc} hosting the TS ML models. 
The use of these components in the testbed enables realistic simulations where UE and cells regularly report KPIs to the RIC and receive real-time reallocation handover requests based on TS handover decisions.
We assess \attackname's impact by simulating two scenarios:
(1) a normal benign scenario, and
(2) a malicious scenario where one cell within the environment executes an attack.
The results of our experiments in the testbed show that on average an adversarial cell was able to obtain a 248.5\% greater UE allocation than it was supposed to in a benign scenario. 
To evaluate our proposed detection method, we test it on simulated malicious scenarios aiming to classify cells' KPIs reports as trusted or untrusted, to identify malicious activity.
\detectname detection method successfully identified malicious cell activity in the test scenarios achieving an accuracy of 99.2\% and an F1 score of 0.978.

The main contributions of this paper can be summarized as follows:
\begin{enumerate}
    \item Present a threat model that takes into account the vulnerabilities resulting from O-RAN's openness and multi-operator untrusted structure.
    We demonstrate an attack (\attackname) in which a cell is the threat actor targeting the TS flow to obtain more UE to serve.
    \item Publish an open-source testbed environment that includes a closed-loop wireless network simulator connected to the OSC RIC cluster that enables realistic and live data collection. 
    \item We also present \detectname; a practical AI-based detection method to mitigate this attack and future attacks based on this threat model.
\end{enumerate}

\section{Background}\label{sec:back}

\subsection{O-RAN Architecture}

Traditional RAN components are monolithic, vendor-provided black boxes that integrate all layers of the cellular protocol stack. 
This design limits reconfigurability, hinders coordination between network nodes, and locks operators into specific vendors. 
In addition, the vendor develops all the components, which drives up costs for network operators and creates vendor lock-in, which limits flexibility and innovation.
To address these challenges, O-RAN has emerged as a new paradigm that leverages disaggregated, virtualized, and software-based components connected through open, standardized interfaces. 
This approach enhances flexibility, supports multi-vendor ecosystems, and enables intelligent, data-driven closed-loop control. 
By adopting cloud-native principles, O-RAN improves RAN resiliency, adaptability, and innovation potential.

\begin{figure}
    \centering
    \includegraphics[width=0.9\columnwidth]{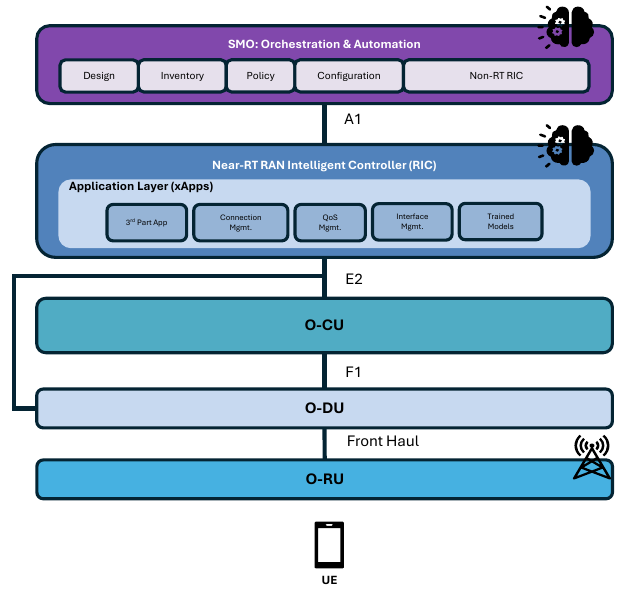 }
    \caption{O-RAN architecture high-level overview.}
    \label{fig:oran-arch}
\end{figure}

The O-RAN specifications build on the 3GPP long-term evolution (LTE) and new radio (NR) standards, extending the 3GPP NR 7.2 functional split for base stations~\cite{polese2023understanding}.
This split disaggregates base station functions into three distinct components: the central unit (CU), distributed unit (DU), and radio unit (RU), enabling greater flexibility and modularity.
These units connect to RICs via open interfaces, enabling the streaming of RAN telemetry and the deployment of control actions and policies.
The O-RAN architecture's components and their open interfaces are illustrated in~\Cref{fig:oran-arch}.

\subsection{The RAN Intelligent Controllers RIC}

The RIC is a key element in the O-RAN architecture, introducing programmable components capable of executing optimization and intelligent routines with closed-loop control.
RICs often leverage AI and ML models for various tasks like network slicing, handovers, and scheduling policies.
By doing so, RICs significantly enhance network efficiency and performance optimization.
Some of the main key benefits of RICs include:
Improved network performance and efficiency through AI and ML-driven optimization~\cite{balasubramanian2021ric}.
Increased flexibility and programmability of the RAN~\cite{chen2023flexslice}.
Reduced operational costs through automation and intelligent resource management~\cite{eugina2024ric}.
Improved UE experience through more granular and intelligent control of network resources~\cite{nisar2023}.
The RIC as described in the O-RAN specifications consists of two primary levels: the near-real-time RIC (near-RT RIC) and the non-real-time RIC (non-RT RIC)~\cite{WG2, WG3}.

The \textbf{Near RT RIC} acts as the near-real-time decision-making core of the network.
It operates with control loop periodicities ranging from 10 milliseconds to 1 second, enabling it to manage dynamic adjustments within the network in near real-time.
Designed to oversee multiple RAN nodes (DUs or CUs), the near-RT RIC can influence the quality of service (QoS) for hundreds or even thousands of UE connections.
The near-RT RIC hosts specialized applications known as xApps, which perform essential tasks such as RAN data analysis, traffic steering, and network control.
These xApps communicate with the RAN via the E2 interface, receiving real-time data and sending control commands back to adjust network behavior. 
Additionally, the near-RT RIC provides APIs and services to support the automated lifecycle management of xApps, including onboarding, deployment, and termination.
These capabilities ensure seamless internal messaging, conflict mitigation, and operational stability across the network~\cite{WG2}.

\textbf{Non-RT RIC} 
The non-RT RIC is responsible for longer-term (operating on timescales greater than one second) network optimization and policy management to support the near-RT RIC via the A1 interface (e.g., the threshold for UE reallocation, in the TS xApp). 
It hosts applications called rApps, which can handle tasks such as network slicing, energy saving, and AI/ML model training~\cite{WG3}.

\subsection{Traffic Steering (TS)}\label{subsec:ts}
\begin{figure}
\includegraphics[width=1\linewidth]{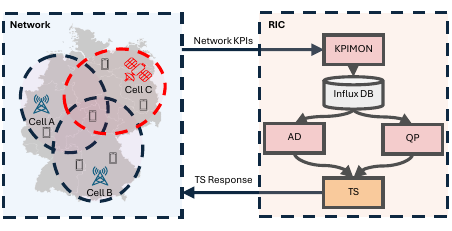}
  \caption{High-level overview of the relevant components in the TS flow.}
  
  \label{fig:ts_flow}
\end{figure}

One of the most important tasks of the near-RT-RIC is the TS task, as it is responsible for managing the UE cells' connection in the network. 
The TS flow illustrated in~\Cref{fig:ts_flow} demonstrates a network topology where UE is located within the reception range of cells A, B, and C and can be connected by each of these cells. 
The network operator needs to make a decision regarding which cell the UE should be connected to.
There are several TS approaches to make this decision. 
Many handovers for the TS task are performed using reinforcement learning (RL), due to its policy-based decision-making nature~\cite{priscoli2020traffic, orhan2021connection}. 
Another way to perform the TS task is to connect UE to a cell based on the maximum received signal reference power (RSRP). 
As a UE moves away from its serving cell, the RSRP from that cell decreases over time, while the RSRP from a nearby target cell increases as the UE approaches it~\cite{tayyab2019survey}. 
One of the most common approaches presented in the OSC, is the QoE prediction, which is based on QoE prediction for potential new target cells (this is the approach we followed in this paper)~\cite{dryjanski2021toward}.
In this TS approach, there are four main xApps involved: KPI monitoring (KPIMON), anomaly detection (AD), QP, and TS.
The TS flow illustrated in~\Cref{fig:ts_flow} contains the following steps;
The KPIMON receives telemetry from the cells regarding the network status (e.g., cells and UE's KPIs) and writes them into the RIC influx database (DB). 
The \textit{AD} xApp, scheduled to run every $10ms$, identifies UE with an anomalous QoE that might need reassignment to another cell. 
Detection is performed by a pretrained isolation forest (iForest) model, based on UE metrics extracted by the KPIMON xApp and stored in the RIC. 
The AD xApp sends this list of anomalous UEs to the \textit{TS} xApp for reallocation.
Then, the TS calls the \textit{QP} xApp for QoE prediction for each neighbor cell of all UE in the list.
The QP trains a vector autoregression (VAR) model for every UE neighbor cell, forecasts the QoE, and sends it back to the TS.
Finally, based on the QP predictions and a given A1 policy from the non-RT-RIC, for each UE, the TS decides whether it should stay in its current serving cell or be handed over to a new target cell.

\subsection{Multiple Operators Deployments}

The evolution of telecommunications infrastructure has witnessed a significant shift toward disaggregation, where different network elements are operated by distinct entities (such as different companies). 
While multi-vendor deployments have been a longstanding practice in telecommunications, the emergence of multiple operators managing different network segments introduces new trust considerations. 
This paradigm shift predating O-RAN was primarily driven by cost-efficiency considerations and the understanding that specialized capabilities could be better managed through outsourcing arrangements~\cite{nec2013ran,farhat2017radio,markendahl2013shared,markendahl2013network,opadere2019energy}.

The infrastructure-sharing model gained particular prominence with the rise of tower companies like American Tower\footnote{American Tower: \url{https://www.americantower.com/why/history}} and Vantage Towers\footnote{Vantage Towers: \url{https://www.vantagetowers.com/en}}. 
These companies emerged as spin-offs from traditional telecom operators and have since evolved into independent entities managing critical network infrastructure. 
The participation of third-party mobile tower companies has also led to an increased incidence of base stations (BS) colocation~\cite{opadere2019energy}. 
For example, in the U.S., companies sublease space from independent landlords to deploy BSs belonging to more than one operator on the same premises~\cite{towergenius2018}.
In the United Kingdom, Freshwave deployment enables four mobile operators to share the same indoor small cell infrastructure, demonstrating the feasibility of extensive operator collaboration~\cite{keith2024oran, wooden2024oran}.
In Germany, Deutsche Telekom and Telefónica Deutschland established a major infrastructure-sharing agreement, in which thousands of mobile sites were covered through reciprocal network access~\cite{dtdf2021oran}. 
Similarly, Vodafone and Orange have implemented extensive network sharing in Spain, particularly focusing on rural and suburban areas~\cite{vodo2019oran}.
Recent developments have also facilitated innovative deployment scenarios. 
For example, the integration of satellite networks as RU/DU units~\cite{esa2024oran} with cellular backbones represents an extreme case of a multi-operator scenario, highlighting both the potential and challenges of such arrangements~\cite{zhu2021integrated, liu2024democratizing}. 
Companies like AST SpaceMobile\footnote{AST:  \url{https://ast-science.com/spacemobile-network/}}, Lynk Global\footnote{Lynk Global: \url{https://lynk.world/}} and Starlink~\cite{liu2024democratizing} are developing satellite-to-cellular solutions that will integrate with terrestrial networks, creating new multi-operator paradigms.

\section{Related Work}

\begin{table*}[!ht]
\small  
\setlength{\tabcolsep}{4pt}  
\centering
\caption{Summary of Related Work}
\begin{tabular}{>{\raggedright\centering\arraybackslash}p{1.5cm}|p{10.5cm}|>{\centering\arraybackslash}p{0.8cm}|>{\centering\arraybackslash}p{1.5cm}|>{\centering\arraybackslash}p{2cm}}
\toprule
\textbf{Paper} & \textbf{Key Points} & \textbf{Actor} & \textbf{Attack} & \textbf{Mitigation}
\\ \midrule
\cite{habler2022adversarial} & A novel AML threat assessment methodology with practical demonstrations and tools for high-risk threats in ML-based network management & UE & Evasion \& Poison & - \\ 
\midrule
\cite{tsourdinis2024ai} & Anomaly traffic detector that enhances network security by mitigating DoS attack executed by UE on RIC xApps & UE & DoS & ML-Based \\ \midrule
\cite{alimohammadikpi} & Presents how compromised KPIs can poison the RIC closed loop followed by an LSTM detection method & - & Poison & LSTM \\ 
\midrule
\cite{soleymaniddos} & 
ML-based detection for preventing DDoS attacks executed by malicious UE
& UE & DDoS & ML-Based \\ 
\midrule
\cite{branco2024evaluation} & 
Fast ML-based detection for DDoS attacks on O-RAN 
& UE & DDoS & ML-Based \\ 
\midrule
\cite{groen2024implementing} & 
MiTM attack that exploits the open interfaces and poisons the network slicing xApp
& - & Poison & DRL AE \\ 
\midrule
\cite{sapavath2023experimental} & 
Presents how a malicious xApp reduces the network capacity by performing FGSM and PGD attacks
& xApp & Evasion & - \\ \midrule
\cite{balakrishnan2024enhancing} &
Evasion attack on the connection management xApp, with defense approaches
& xApp & Evasion & Adv Training \\ 
\midrule
\textbf{Our Paper} & Attack by untrusted cell in O-RAN with AI-driven detection & Cell & Evasion & LSTM-AE \\ \bottomrule
\end{tabular}
\label{tab:related-work}
\end{table*}

\subsection{O-RAN Security}\label{sec:sec-oran}

New concepts and technologies are continually being introduced into the RAN, each bringing new cybersecurity threats and significantly expanding the RAN’s attack surface~\cite{soltani2024intelligent, parvez2018survey,polese2023understanding, ahmad2018overview,azariah2024survey}.
O-RAN security focuses on several key areas. 
Recent attacks on traditional RANs are reviewed to assess their applicability to the O-RAN architecture~\cite{shen2022security}. 
The open-source and disaggregated nature of O-RAN introduces unique threats, necessitating analysis of the security implications of the O-RAN architecture~\cite{mimran2022evaluating}. 
Specific vulnerabilities arise from O-RAN’s openness, particularly in xApp access control and the E2 interface, which could allow unauthorized access and manipulation of network policies~\cite{hung2024security}. 
The classification of various security-related risks specific to O-RAN includes inadequate logging, lack of encryption, and insufficient access controls, which can lead to security breaches and data integrity problems~\cite{liyanage2023open}.
The comprehensive summary of the security threats, requirements, and recommended mitigation strategies associated with the O-RAN framework provided by Park et al.~\cite{park2024investigation} highlights the steps required to strengthen the architecture’s resilience against emerging threats.
However, as emphasized by Park et al.~\cite{park2024investigation}, there are many remaining security risks, as we will demonstrate in this paper. 

\subsection{Attacks on O-RAN}
Adversaries may exploit the inherent vulnerabilities of learning algorithms, and specifically ML algorithms, with various attack techniques, which are referred to as adversarial machine learning (AML)~\cite{balakrishnan2024enhancing, habler2022adversarial,sapavath2023experimental,baguer2024attacking}.
Recent work~\cite{habler2022adversarial} provided a comprehensive threat assessment of ML usecases within O-RAN according to a common cybersecurity risk assessment (NIST ontology).
In their work, the authors outline the potential adversaries, their capabilities, and their goals, identify threats to ML production systems within O-RAN, and enumerate attacks that can materialize these threats. 
In addition to their threat modeling, the authors demonstrated how UE can produce manipulated signals that lead to incorrect anomaly detection and QoE classification. 
By doing so, the UE influences the model's decision-making process by presenting inputs that are very similar to legitimate data but designed to cause misclassification or incorrect predictions~\cite{habler2022adversarial}. 

Sapavath et al.~\cite{sapavath2023experimental} conducted experiments in an O-RAN testbed to demonstrate that both the fast gradient sign method (FGSM) and projected gradient descent (PGD) attacks can effectively manipulate input data for the xApp interference classifier, leading to misclassification. 
Even minimal adversarial perturbations (i.e. small modifications to input data designed to mislead ML models) have been shown to drastically impair the xApp’s accuracy, which consequently reduces network capacity and increases overall bit loss within the O-RAN system~\cite{sapavath2023experimental}.

The authors of~\cite{baguer2024attacking} analyze O-RAN WG11’s~\cite{WG11} threat model and risk assessment methodology, focusing on DoS and performance degradation threats. 
They identify specific vulnerabilities, mapping them to potential attacks across critical O-RAN interfaces.
Through experiments using an O-RAN deployment, the authors evaluated the impact of DoS and performance degradation attacks on these interfaces, assessing their resilience in various attack scenarios~\cite{baguer2024attacking}.

\subsection{O-RAN Attack Mitigation}

Growing concerns over attacks on O-RAN ML systems have prompted research on mitigation of such attacks.

Both white-box and black-box evasion attacks have been demonstrated on deep reinforcement learning (DRL)-based traffic steering~\cite{orhan2021connection}, revealing the impact of adding noise into the UE's metrics. 
These attacks showed that PGD attack can reduce coverage rates by as much as 50\%, while jamming attacks result in up to a 25\% reduction in coverage.
To counter these effects, two mitigation techniques were proposed: (1) adversarial training~\cite{goodfellow2014explaining}, and (2) regularized training~\cite{zhang2020robust}. 
Both techniques were found to improve model robustness against such attacks~\cite{balakrishnan2024enhancing}.

Recent work by Xavier et al.~\cite{xavier2023machine} introduced an effective mitigation strategy for several types of DoS attacks. 
This approach employs ML models to analyze air interface measurements, enabling the early detection of malicious traffic before it disrupts network services.
In their followup work, the authors improved their mitigation strategy which is supposed to mitigate all types of DoS attacks while improving its ability to be deployed on real systems~\cite{xavier2024cross}.

Research directly related to our study on attacks executed by malicious KPIs reaching the RIC has also been performed. 
In~\cite{groen2024implementing}, the authors highlighted the vulnerabilities introduced to the intelligent components of O-RAN due to the adoption of open interfaces. 
Man-in-the-middle attack (MiTM) attackers were shown to be able to inject malicious KPI reports into the E2 interface targeting the near-RT RIC or deliver malicious control actions from the near-RT RIC to E2 nodes.
They specifically demonstrated AML attacks on the input KPI reports of a network slicing xApp. 
To mitigate such threats, the authors proposed a method based on AEs to detect this threat~\cite{groen2024implementing}. 
Another study~\cite{alimohammadikpi} built on this type of poisoning attack, emphasizing the critical role of KPIs in near-RT RIC control loop use cases. 
The authors proposed an LSTM model for detecting anomalous KPIs, which was shown to be a robust approach for mitigating such attacks in their evaluation.

To detect Distributed DoS (DDoS) attacks et al~\cite{branco2024evaluation} demonstrates that XGBoost achieves high precision and recall with the fastest execution time compared to random forest and multilayer perceptron, ensuring operations remain within latency requirements.
Another study~\cite{soleymaniddos} presents a ML-based framework for detecting DDoS attacks in O-RAN, utilizing dApp and xApps to enhance real-time threat detection, while balancing speed and accuracy. 
It evaluates multiple ML algorithms to identify the best-fit models for anomaly detection and service usage tracking, addressing O-RAN's unique challenges.

To the best of our knowledge, no recent research has considered network cells as untrusted elements in the RAN as potential threat actors or proposed mitigation strategies for such types of attacks.
The related works~\cite{balakrishnan2024enhancing, sapavath2023experimental, groen2024implementing, branco2024evaluation, soleymaniddos, alimohammadikpi, habler2022adversarial} performed on O-RAN attacks and mitigation is summarized in~\Cref{tab:related-work}.

\section{Threat Model}\label{sec:tm}
We present a threat model based on the NIST ontology for modeling an enterprise
security~\cite{tabassi2019taxonomy, bitton2023evaluating}. 
The proposed threat model considers the following assumptions:
\\
(1) \textbf{Multiple Operators RAN Deployments.}
The shift of telecommunications infrastructure to disaggregation resulted in different network elements being operated by distinct entities, allowing the reduction of operational cost, achieving both capital expenditures (CAPEX) and operational expenditures (OPEX) savings~\cite{nec2013ran,farhat2017radio,markendahl2013shared,markendahl2013network,opadere2019energy}. 
Implementation reports and real-world scenarios to support the feasibility of multi-operator deployment are detailed in the background section (\cref{sec:back}).
In addition, as described in 3GPP specifications, multiple cell operators agree on sharing a coverage area taking into account the load balancing between the cells\cite{3gpp2013ran}.
Similarly, GSMA also released a specification on infrastructure sharing, describing standards for site, tower, RAN and core network sharing~\cite{gsma2018}.
\\
(2) \textbf{Financial Model.} 
In multi-operator deployments, when an operator lacks the resources to serve its clients (UEs), services are provided through a third-party operator.
According to Farhat et al.~\cite{farhat2015access,farhat2014best} UE payments go to their home operator, and the latter must pay a service price (transaction cost) to the new access operator.

Under these assumptions, an attacker is a malicious cell operator carrying an AML attack targeting the O-RAN TS flow. 
The attack goal is to gain unfair UEs allocation, thereby increasing its revenue. 
Such manipulation can degrade the QoE for the victims' UEs and reduce the income of its neighboring benign operators.

\begin{figure}[hbt]
    \centering
    \includegraphics[scale=0.35]{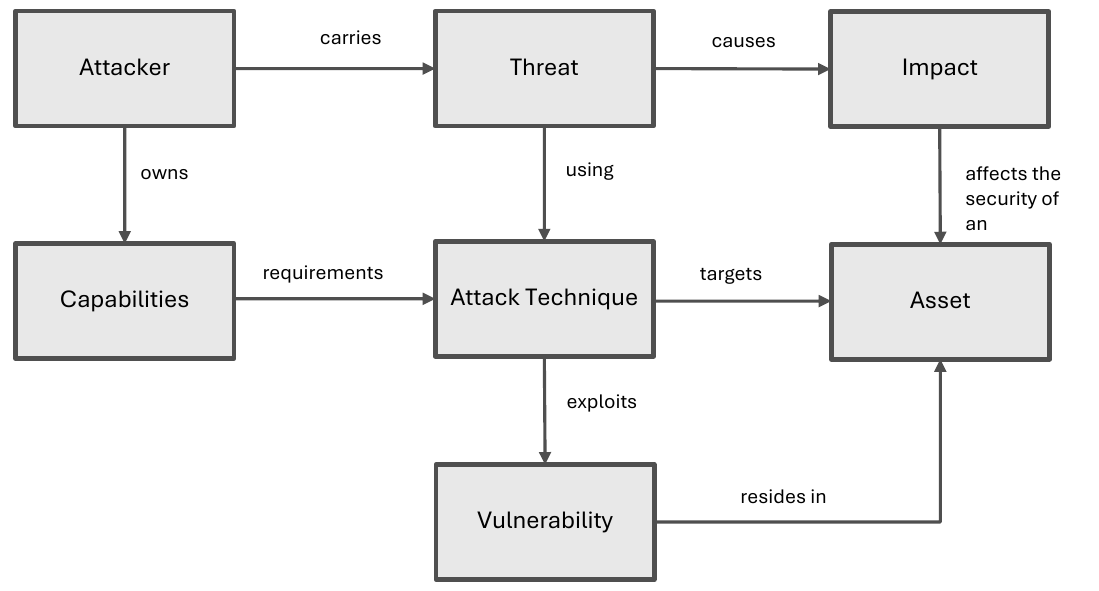}
    \caption{Threat analysis based on NIST ontology~\cite{tabassi2019taxonomy}.}
    \label{fig:nist}
\end{figure}

The threat model entities and their relations based on NIST ontology as illustrated in~\Cref{fig:nist} are described as follows:
\begin{itemize}
    \item \textbf{Attacker:} A malicious operator running a malicious cell operating in the O-RAN network. 
    In the remainder of the paper, we will use the term malicious cell.
    \item \textbf{Adversarial Capabilities:}
    (a) The malicious cell can manipulate the KPIs it reports to the RIC.
    (b) The malicious cell has knowledge of the targeted TS task flow. 
        This capability follows SOTA AML threat analysis in O-RAN~\cite{habler2022adversarial}.
    \item \textbf{Threat}: A malicious cell disrupts the TS process by influencing the QP model's QoE predictions, resulting in an unfair allocation of UE to the malicious cell.
    \item \textbf{Vulnerability:} Refers to the inherent ability to manipulate the input of the ML model used by the QP, causing it to inaccurately predict QoE.
    \item \textbf{Technique:} Query-based evasion attack.
    \item \textbf{Assets:} The TS task hosted on the near-RT RIC, responsible for the allocation of UE to cells.
    \item \textbf{Impact:} The attacker is able to serve more UE than it is supposed to, which can:
    (a) malicious cell to receive payment for providing the service to UEs that would gain better service from other cells.
    (b) reduce the QoE of UE affected by the attack, since providing service to UE that would receive a better QoE by other cells.
\end{itemize}

\section{\attackname Attack}\label{sec:attack}

\begin{figure*}[!ht]
    \centering
    \includegraphics[width=1\linewidth]{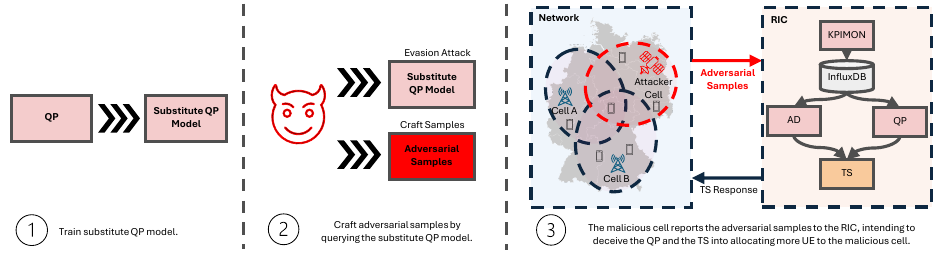}
    \caption{Attack flow steps:
    (1) the attacker trains a substitute QP model, replicating the behavior of the target model;
    (2) the attacker employs an adversarial evasion attack to generate adversarial samples; and 
    (3) the attacker reports its generated adversarial samples to mislead the TS to allocate more UE.}
    \label{fig:attack}
\end{figure*}

The \attackname attack (adversarial perturbation against traffic efficiency), an attack designed to manipulate network TS by attacking the QP model is illustrated in~\Cref{fig:attack}. 
The attacker's objective is to maximize the amount of UE assigned to attacker's service.
To achieve this, the attacker manipulates its own KPIs to mislead the QP model into forecasting a higher QoE than the actual QoE.
To execute this attack, the malicious cell needs to know which TS approach (see~\Cref{subsec:ts}) is implemented on the network; this will allow the attacker to replicate the behavior of the target model.

The attack unfolds in three main stages:
(1) The attacker begins by training a substitute QP model, replicating the behavior of the target model used in the TS flow.
(2) Using the substitute model,  the attacker learns the model’s decision boundaries and behavior. 
By analyzing the decision boundaries, the attacker identifies the minimal input perturbations that need to be added to the cell's KPIs to manipulate the QP model.
Then the attacker performs an adversarial evasion attack to generate adversarial samples.
(3) The crafted adversarial samples, representing the perturbed KPIs, are reported to the RIC as legitimate data.
These adversarial samples are written to the RIC database by the KPIMON xApp. 
When the AD xApp detects that UE might require cell allocation, the TS xApp requests the QP xApp for a QoE prediction for the potential new target cells (as detailed in~\Cref{subsec:ts}). 
At this point, the QP model predicts an artificially high QoE for the malicious cell based on the adversarial samples, causing the TS xApp to allocate UE to the malicious cell.

The proposed attack is formally described as follows: 
let $N=(V^{cl},V^{ue},E)$ denote the network bipartite graph where $V^{cl}=\{v^{cl}_1, v^{cl}_2, ..., v^{cl}_n\}$ are the cell's nodes and $V^{ue}=\{v^{ue}_1, v^{ue}_2, ..., v^{ue}_m\}$ are the UE's nodes, and $n$ and $m$ are respectively the number of cells and number of user equipments. 
The edges $E\subseteq V^{cl}\times V^{ue}$ are defined by serving connections between pairs of cells and UE. For example, if $v^{cl}_x$ is the serving cell of UE $v^{ue}_y$, then $(v^{cl}_x,v^{ue}_y) \in E$.

The objective of the attacker cell $v_{adv}^{cl}$ is to find a perturbation noise $\delta$ that can be added to its KPI reports $R$, such that the QP model will predict a higher QoE than it should as presented in~\cref{eq:attack} where $\delta^{*}$ represents the optimal perturbation, $\mathcal{Q}$ is the QP model, and $y$ is the true QoE prediction.
The attacker sends these crafted adversarial samples to the RIC where they are incorporated into the TS flow, as detailed in~\cref{subsec:ts}. 
This manipulation leads the QP model to overestimate the QoE for the adversarial samples, potentially resulting in unjustified greater UE allocations $(v_{adv}^{cl} ,v^{ue})$ to the adversarial cell.

\begin{subequations}\label{eq:attack}
\centering
\begin{align}
R_{adv} = R + \delta^{*}
\label{eq:attack:obj} \\
\delta^* =\arg\max_\delta(\mathcal{L}(\mathcal{Q}(R+\delta), y)) 
\label{eq:attack:delta} 
\end{align}
\end{subequations}

\section{\detectname Detection Method}\label{sec:frame}
\begin{figure}[!hb]
\centering
   \includegraphics[width=1\linewidth,]{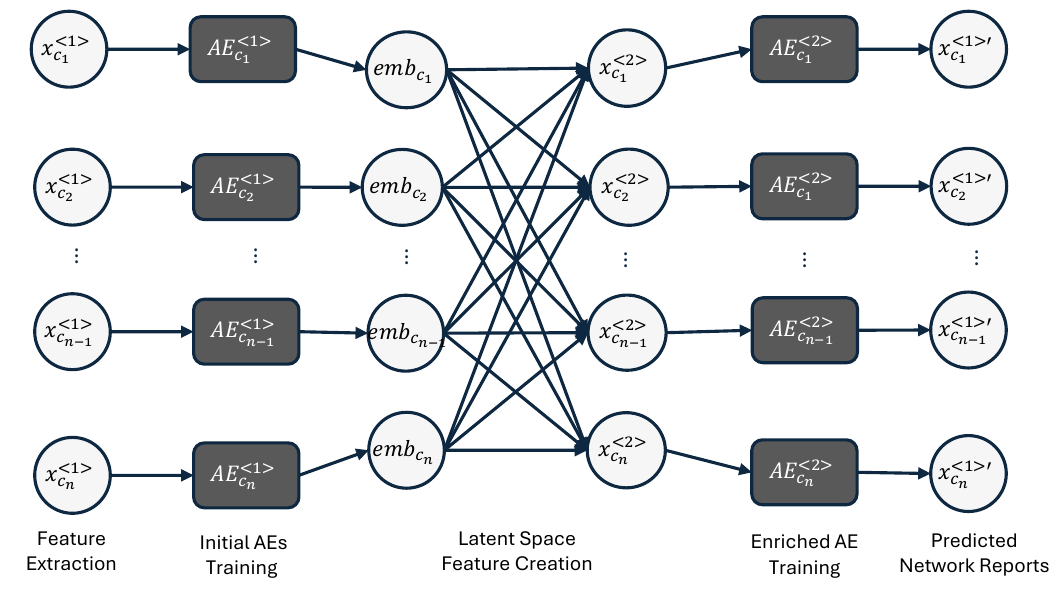}
    \caption{LSTM-autoencoder framework architecture.}
  \label{fig:AE-LSTM}
\end{figure}
The \attackname attack poses a significant threat to resource management by targeting the network TS. 
To mitigate these types of attacks, we propose \detectname (monitoring adversarial RAN reports)—a framework designed to detect adversarial cell telemetry (reported KPIs) in real time. 
\detectname can be deployed as an xApp on the near-RT RIC, providing immediate notifications about whether a cell's telemetry can be trusted.
The detection architecture is based on LSTM networks, which have proven to be particularly effective for anomaly detection in time-series data, especially when combined with AE architectures~\cite{lee2024lstm, sagheer2019unsupervised}. 
LSTM AEs are trained to reconstruct normal input sequences, leveraging an LSTM's ability to capture long-term dependencies in time-series data~\cite{wei2023lstm}.
We utilize a two-layer combination of LSTM and AEs to capture not only individual cell behavior but the contextual interactions of a specific cell within the entire network.

The detection framework architecture, which is illustrated in \cref{fig:AE-LSTM}, operates as follows:
(1) Feature Extraction and Initial AE Training: 
Relevant time-series features are extracted from the KPIs reported by cells and UE. 
Then, for every cell, a dedicated AE model is trained to learn the patterns in these features.
(2) Latent Space Feature Creation: 
Using the latent representations from the first set of AE models, new enriched feature vectors are generated. 
This is achieved by concatenating the latent space of a specific cell's AE with aggregated latent spaces from other cells, effectively capturing both local and global network contexts.
(3) Enriched AE Training:
A second AE is trained for each cell, however this time using the enriched feature vectors as input.
The objective is to reconstruct the original features from these vectors, leveraging the additional contextual information.
(4) Classification of Network Reports:
Finally, a classifier is trained to compare the reconstructed output of the second AE with the original input features from the first AE. 
If the reconstruction loss exceeds a predefined threshold, the input is labeled as untrusted; otherwise, it s labeled as trusted.
By integrating contextual insights and multi-stage reconstruction, \detectname serves as a robust mechanism for the detection and mitigation of adversarial activity in the network.

Formally MARRS is described as follows: 
Given a network denoted as in~\Cref{sec:attack},
and the network KPI reports $R$,
we aim to identify a framework $\mathcal{F}$ that will reconstruct the original network reports.
We examine the loss function score
$\ell(\mathcal{F}(R), R)$ between the reconstructed report $\mathcal{F}(R)$ and the original report $(R)$.
If it is higher than the threshold $\mathcal{T}$, the classifier $\mathcal{C}$ will return trusted (i.e., 0), otherwise, untrusted (see~\Cref{eq:clas}).

\begin{equation}\label{eq:clas} 
\mathcal{C}(R,\mathcal{T}) = 
     \begin{cases}
       \text{0,} &\quad\text{if,   $\mathcal{T}$} > \ell(\mathcal{F}(R), R)\\
       \text{1,} &\quad\text{if,   $\mathcal{T}$}\le\ell(\mathcal{F}(R),R)\\
     \end{cases}
\end{equation}

\subsection{Feature Extraction and Initial AE Training}
\subsubsection{Feature Extraction}
 
We begin by extracting relevant time-series features from the network, summarized in~\Cref{tab:x1}. 
For a given O-RAN network, we extract KPIs such as the physical resource block (PRB) aggregation period, PRB downlink/uplink ratios, and the amount of UE currently in the cell, newly entering UE, and UE leaving the cell.
Additionally, we extract UE-specific metrics, including the average and standard deviation of the Packet Data Convergence Protocol (PDCP) downlink and uplink throughput, UE PRB downlink/uplink ratios, and signal quality indicators such as the reference signal received power (RSRP) and signal-to-noise ratio (RSSNIR). 
In total, these features result in 11 time-series features for each cell. 
These features were extracted as they are the ones used in the TS flow (defined in the OSC RIC~\cite{bimo2022osc}).
Finally, all features are standardized to produce scaled values and split to sliding time windows denoted as $X_{c_i}^{<1>}$.

\begin{table*}[!ht]
\small
\centering
\setlength{\extrarowheight}{-3pt}
\caption{Cell and UE Feature $X^{<1>}$}
\begin{tabular}{
    p{0.04\textwidth}|
    p{0.13\textwidth}|
    p{0.04\textwidth}|
    p{0.71\textwidth}
}
\toprule
\textbf{Type} & \textbf{Feature Name} & \textbf{Units} & \textbf{Description} \\ 
\midrule
\multirow{5}{*}[-0.3em]{\rotatebox[origin=c]{90}{Cell KPIs}} 
& Throughput & bps & The amount of data transmitted per unit of time across a cell \\[-2pt]
\cmidrule{2-4}
& MeasPeriodPrb & kHz & Physical Resource Block (PRB) is defined as a time-frequency resource in the physical layer of wireless communication systems \\[-2pt]
\cmidrule{2-4}
& Number\_UEs & \# & Amount of UEs the cell is currently serving \\[-2pt]
\cmidrule{2-4}
& New\_UEs & \# & Amount of new UEs in the cell \\[-2pt]
\cmidrule{2-4}
& Left\_UEs & \# & Amount of UEs that left the cell \\[-2pt]
\midrule
\multirow{6}{*}[-0.2em]{\rotatebox[origin=c]{90}{\shortstack{Aggregated\\ 
UE KPIs}}} 
& ThpDl\_Mean & \multirow{2}{*}{bps} & \multirow{2}{*}{UE's downlink throughput} \\
& ThpDl\_Std & & \\[-2pt]

\cmidrule{2-4}
& Rssnir\_Mean 
& \multirow{2}{*}[-1pt] {dB}
& \multirow{2}{*}{\shortstack[l]{RSSNIR (signal to interference \& noise ratio) The ratio of the useful signal power to the\\combined interference and noise power}} \\ [2pt]
& Rssnir\_Std & \\[-2pt]
\cmidrule{2-4}

& Rsrp\_Mean 
& \multirow{2}{*}[-1pt]{dBm} 
& \multirow{2}{*}[-1pt]{RSRP (Reference Signal Received Power) - The signal strength received by UE from cell} \\
& Rsrp\_Std & & \\ 

\bottomrule
\end{tabular}
\label{tab:x1}
\end{table*}

\subsubsection{Initial AE Training}

After extracting $X_{c_i}^{<1>}$ from the network, we proceeded to train a dedicated AE for each cell, denoted as $AE_{c_i}^{<1>}$. 
Each AE is based on an LSTM AE architecture and designed to encode every time window $x_{c_i}^{<1>} \in X_{c_i}^{<1>}$ to a new dimensional latent space representation $emb_{c_i}$ and then decode it back to the original $x_{c_i}^{<1>}$. 
Completing this phase results in a trained LSTM AE for each cell in the network $AE_{c_i}^{<1>}$.

\subsection{Latent Space Feature Creation}

In this phase, we conduct a second round of feature extraction to generate enriched feature vectors for each cell, capturing both contextual information from the specific cell and the entire network. 
To achieve this, we leverage the embedded ($emb_{c_i}$) latent space representations produced by the trained $AE_{c_i}^{<1>}$ from the initial feature set $X_{c_i}^{<1>}$.
For each cell $v^{cl}_i$, we construct a new feature set by concatenating its latent space embedding with an aggregated embedding (e.g.,  average) derived from the rest of the network. 

\begin{equation}\label{eq:x2}
  X_{c_i}^{<2>} = {emb_{c_i}}^ \frown {\frac{1}{n-1}\sum_{j\in V^{cl}\setminus{\{v^{cl}_i}\} } emd_{c_j}}  
\end{equation}
where $c_i$ and $c_j$ represent the features associated with the cell nodes $v^{cl}_i$ and $v^{cl}_j$ respectively and $n = |V^{cl}|$ represents the number of cells in the network.
This approach enriches the feature set by capturing both local and network-wide contextual information.

\subsection{Enriched AE Training}

The next step in our proposed framework involves training a second round of AE models $AE_{c_i}^{<2>}$ for each cell $v^{cl}_i\in V^{cl}$. 
In this phase, the $AE_{c_i}^{<2>}$ models are designed to encode the enriched feature set $X_{c_i}^{<2>}$ to a new latent space using LSTM layers similar to what was done in $AE_{c_i}^{<1>}$.
However, unlike the initial round, these new AEs are trained not to reconstruct their input ($x_{c_i}^{<2>}$) but to decode and reconstruct the first feature set $x_{c_i}^{<1>}$.
Through this process the AE models learn to incorporate information from the entire network while effectively leveraging the specific reports from the individual cell, resulting in a more network-context-aware representation.

\subsection{Classification of Network Reports}\label{subsec:class}

After completing the second round of AE training, the framework is ready to detect malicious activity. 
To classify and detect this activity, we propagate the cell reports throughout the framework; when cell reports reach the RIC, the first feature set is extracted ($X_{c_i}^{<1>}$) and encoded by its $AE_{c_i}^{<1>}$, producing the latent embedding ($emb_{c_i}$).
This embedding is then used to generate the enriched feature set $X_{c_i}^{<2>}$ as defined in~\Cref{eq:x2}, which is fed into the second AEs noted as $AE_{c_i}^{<2>}$ to reconstruct the first feature set as $X_{c_i}^{<1>'}$. 
If the models are trained effectively, the reconstruction loss $\ell$ between the given feature set extracted from time window report $x_{c_i}^{<1>}$ and the framework output $x_{c_i}^{<1>'}$ is expected to be low for benign reports and high for malicious or compromised reports. 

To classify these reports, a threshold $\mathcal{T}$ needs to be defined based on a certain policy provided by the operator.
The policy should determine which classification metrics (e.g., recall, precision, F1 score) should be optimized depending on the operator's preferences regarding the network's performance.  
Finally, reports with reconstruction loss $\ell$ exceeding $\mathcal{T}$ are classified as untrusted, while those below $\mathcal{T}$ are classified as trusted as in~\Cref{eq:clas}.

\subsection{Sequence-Based Detection (S-\detectname)}\label{subsec:sbd}
We propose an extension to \detectname approach which is based on sequence detection denoted as S-\detectname.
During inference, given a time window of the same size used in training, we expect the framework to yield a low reconstruction loss for benign time windows and significantly higher losses for malicious time windows.
In this approach, both malicious inputs and unrelated outliers that exceed the framework's reconstruction loss threshold are classified as untrusted. 
This can result in a higher false positive rate (FPR) in the detection method, if a benign sample is classified as untrusted (positive) falsely.
To address this issue and reduce the FPR, we apply detection based on sequences of time windows, using specific classification rules defined on the entire sequence.
In this detection method, given a trained framework $\mathcal{F}$, sequence of time windows $S=(R_1, R_2,...,R_k)$, classification rule $\mathcal{RL}$ (e.g. "majority vote"), and threshold $\mathcal{T}$, we classify the entire sequence as either trusted or untrusted according to the rule. 
For example, consider a sequence of network time windows' KPI reports $S=(R_1, R_2,...,R_k)$ size $k$, "majority vote" rule, and classifier $\mathcal{C}$ (as in~\Cref{eq:clas}), we classify reports $(R_1, R_2,...,R_k)$:
\begin{equation}\label{eq:seqexmplae} 
\mathcal{CS}(S,\mathcal{T}) = 
     \begin{cases}
       \text{untrusted,} &\quad\text{if, }    \frac{k+1}{2} \le \sum_{j=1}^{k}{\mathcal{C}(R_j,\mathcal{T})}\\
       \text{trusted,} &\quad\text{else }   \\
     \end{cases}
\end{equation}
In this example, if most of the sequence exceeds $\mathcal{T}$, the entire sequence is classified as untrusted.
This approach allows us to reduce the false positives that may arise due to outliers and focus more on malicious behavior.

\section{Test Environment}\label{sec:env}

\begin{figure}[ht]
    \centering
    \includegraphics[width=1\linewidth]{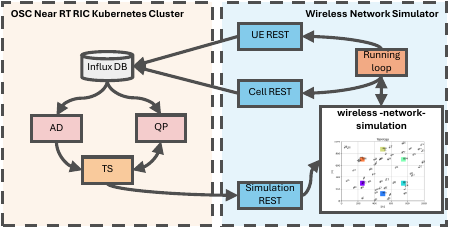}
    \caption{Testbed Environment; left - OSC near-RT RIC Kubernetes cluster and right - the wireless network simulator. }
    \label{fig:testbed}
\end{figure}

We developed a simulation testbed environment that contains two primary components, as illustrated in~\cref{fig:testbed};
(1) Wireless Network Simulator ~\cite{de2022satellite}:  Which simulates real-time network scenarios involving UE and gNB cells.
(2) O-RAN Software Community (OSC) Near-RT RIC Platform~\cite{bimo2022osc}: Deployed within a Kubernetes cluster.
This platform serves as a dynamic hosting environment for the relevant xApps in the TS flow, including the AD, QP, and TS xApps.

In deployment, these two components are isolated from each other to simulate real-world usage and communicate within a closed-loop system via a REST API. 
The interaction occurs as follows:
At the end of each simulation iteration, the simulator reports the current KPIs (both UE and cell KPIs) to the RIC cluster.
Simultaneously, the TS xApp in the RIC platform generates handover requests based on those KPIs and sends them back to the simulator for UE allocation.

\subsection{Wireless Network Simulator} 

The simulator is deployed separately from the near-RT-RIC cluster on an AWS EC2 instance running Ubuntu 20.04 LTS (Focal Fossa). 
The simulation begins with the configuration of network parameters, including the geographical topology size, initial locations and velocities of UE, and cell types (e.g., gNB, eNB) along with their respective locations. 
The simulation operates with two parallel threads, one of which manages the core simulation loop, while the other hosts a Flask app that listens for handover requests from the RIC. 
Upon receiving a request, the Flask app processes it and updates the target cell allocation for the relevant UE.
In each simulation iteration, UE either moves randomly or following predefined trajectories, and the simulator generates updated metrics for both UE and cells, formatted to align with the RIC requirements. 
Each report has the following KPIs: UE ID, serving cell ID, location, timestamp, PDCP aggregation period, PDCP throughput, PRB report timestamp, PRB aggregation period, PRB throughput ratios, reference signal received power (RSRP), reference signal received quality (RSRQ), and signal-to-noise ratio (SNIR). 
Cell KPIs include the cell ID, timestamp, PDCP aggregation period, PDCP throughput, PRB aggregation period, and PRB throughput. 
In this way, the TS handover requests are reflected in real time in the simulation, allowing live UE allocation.

\subsection{OSC RIC Cluster} 
The RIC platform, deployed as a Kubernetes cluster on an AWS EC2 instance running Ubuntu 20.04 LTS (Focal Fossa), hosts the TS flow xApps (AD, QP, and TS) as individual pods. 
It receives KPIs from the simulator, which functions as the RAN. 
Upon receiving these KPIs, the KPIMON xApp processes the metrics and writes them to the RIC’s InfluxDB pod.
Once the data is populated in the RIC database, the TS flow begins as described in~\Cref{subsec:ts}, generating a handover request.
This request is sent back to the simulator, completing one iteration of the closed-loop testbed. 
The process repeats continuously until the simulation loop is done.

\section{Evaluation}\label{sec:eval}

\subsection{Experimental Setting}\label{subsec:settings}

\begin{figure*}[] 
    \centering
    \begin{subfigure}{0.49\textwidth} 
        \centering
        
        \includegraphics[width=\textwidth]{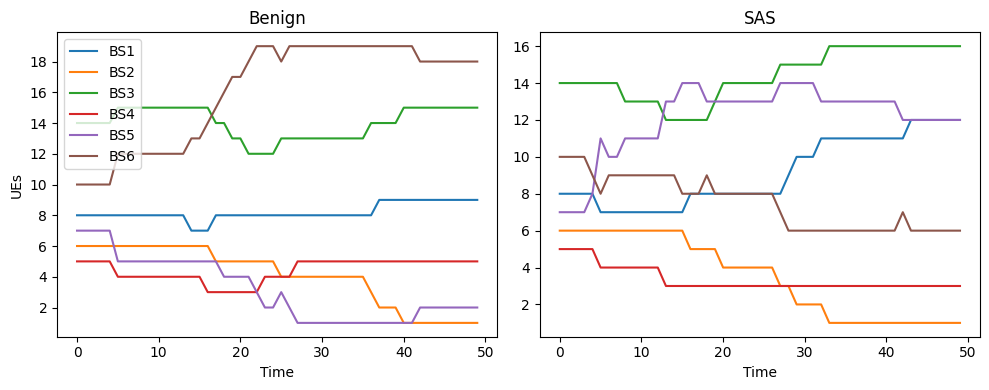} 
        \caption{SAS; BS5 (purple) is the malicious cell.}
        
        \label{fig:attack_scenarios:SAS}
    \end{subfigure}
    \hfill
    \begin{subfigure}{0.49\textwidth}
        \centering
        
        \includegraphics[width=\textwidth]{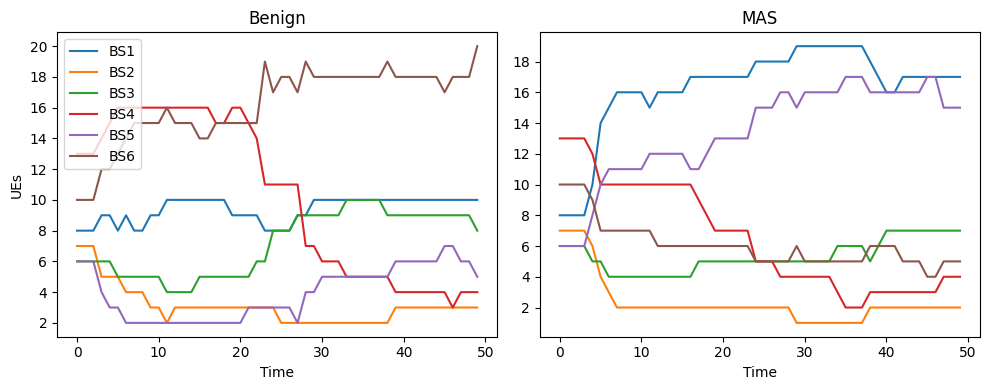}
        \caption{MAS; BS5 (purple) and BS1 (blue) are the malicious cells.}
        \label{fig:attack_scenarios:MAS}
    \end{subfigure}
    \caption{Amount of UE for each cell (BS1-BS6) during each of the network iterations (x-axis) in the examined scenarios: the SAS, the MAS, and the corresponding benign scenario.
    }
    \label{fig:attack_scenarios}
\end{figure*}

\begin{table*}[htbp]
    \small
    \centering
    
         \caption{Amount of UE for each cell in each scenario - corresponding to~\Cref{fig:attack_scenarios}.}
         \label{tab:attacks_sum}
    \begin{tabular}{|c|ccc|ccc|c|ccc|ccc|c|}
        \toprule
        
        & \multicolumn{7}{c|}{SAS} & \multicolumn{7}{c|}{MAS} \\
        \cmidrule{2-15}
        & \multicolumn{3}{c|}{Benign} & \multicolumn{3}{c|}{Malicious} & Difference & \multicolumn{3}{c|}{Benign} & \multicolumn{3}{c|}{Malicious} & Difference \\
        \cmidrule{1-8}\cmidrule{9-15} {cell ID}
        & mean & min & max & mean & min & max & \% & mean & min & max & mean & min & max & \% \\
        \midrule
        BS1 & 
        8.14 & 7 & 9 & 8.81 & 7 & 12 & 108.25\% & 
        8.97 & 8 & 10 & 13.83 & 8 & 19 & 154.16\% \\
        BS2 & 
        4.64 & 1 & 6 & 4.21 & 1 & 6 & 90.77\% & 
        4.27 & 2 & 7 & 3.73 & 1 & 7 & 87.46\% \\
        BS3 & 
        14.01 & 12 & 15 & 14.31 & 12 & 16 & 102.14\% & 
        6.82 & 4 & 10 & 5.55 & 4 & 7 & 81.40\% \\
        BS4 & 
        4.59 & 3 & 5 & 3.83 & 3 & 5 & 83.49\% & 
        11.04 & 3 & 16 & 8.45 & 2 & 13 & 76.53\% \\
        BS5 & 
        4.27 & 1 & 7 & 10.61 & 7 & 14 & 248.50\% & 
        4.56 & 2 & 7 & 11.21 & 6 & 17 & 245.68\% \\
        BS6 & 
        14.34 & 10 & 19 & 8.21 & 6 & 10 & 57.27\% & 
        14.34 & 10 & 20 & 7.23 & 4 & 10 & 50.39\% \\
        \bottomrule
    \end{tabular}
\end{table*}

\subsubsection{Attack} 

\textbf{Data Collection}
To demonstrate and evaluate the impact of the \attackname attack, we set up a network topology with six gNB cells and 50 UEs randomly moving within this topology as shown in~\Cref{fig:bening}. 
At the end of each iteration in the simulation loop, the simulator reports the state of the network to the RIC cluster as described in~\Cref{sec:env}.
Then, the KPIMON xApp populates the RIC DB with the reported KPIs, and the TS flow begins.
Once the TS makes a handover decision (as described in~\Cref{sec:tm}) for the UE, the TS sends it back to the simulator.  
The simulator receives the TS handover request and updates the environment based on its decision. 
\\
\textbf{Attack Scenarios}
We evaluate \attackname in two attack scenarios using the described closed-loop simulation:
(1) Single-Attack Scenario (SAS): A single malicious cell (BS5) executes the \attackname attack, manipulating its KPI reports to mislead the TS into allocating it more UEs.
(2) Multi-Attack Scenario (MAS): Two malicious cells (BS1 and BS5) simultaneously execute the \attackname attack, manipulating their KPI reports to mislead the TS into allocating them more UEs.

For comparison, we establish corresponding benign baseline scenarios where all cells report trusted KPI telemetry to the RIC. 
To accurately model real-world attack progression, we initialize both attack scenarios using identical conditions to their benign baseline scenarios, while the benign scenarios initialized randomly. 
The velocity steps are consistent across all scenarios.
\\
\textbf{Adversarial Sample Generation}
To execute the \attackname, we employed the HopSkipJump attack~\cite{chen2020hopskipjumpattack} from the Adversarial Robustness Toolbox (ART)~\cite{nicolae2018adversarial} to generate adversarial samples. 
This involved categorizing the QP outputs into four quality levels: poor, average, good, and excellent, with the goal of manipulating the QP to forecast a higher quality category to the attacker cell than the true one.

\subsubsection{Detection}

\textbf{Data Collection} To evaluate \detectname's detection capabilities, we train the framework as described in~\cref{sec:frame}.
The training process begins with data collection, conducted through closed-loop benign simulation scenarios. 
In these scenarios, we initialize a network topology consisting of six gNB cells and 50 UEs, randomly moving within the topology as illustrated in~\Cref{fig:bening}. 
From these simulations, we extract relevant features, as detailed in~\Cref{tab:x1} resulting overall dataset for a training size of 10531 records. 
\\
\textbf{Model Training}
The feature set, denoted as $X^{<1>}$, serves as the input for training the first layer of $AE^{<1>}$s in the framework.
The $AE^{<1>}$ architecture consists of an LSTM encoder followed by an LSTM decoder with a fully connected (FC) output layer. 
The models are implemented in PyTorch~\cite{paszke2017automatic}, using the Adam optimizer and mean squared error (MSE) as the loss function. 
After training the $AE^{<1>}$s, we extract the second feature set ($X^{<2>}$), according to~\cref{eq:x2}, which is subsequently used to train the next layer of AEs ($AE^{<2>}$).
The $AE^{<2>}$s use the same architecture, optimizer, and loss function as the $AE^{<1>}$s. 
Both AE layers are trained for 200 epochs and the hyperparameters such as the number of LSTM layers hidden size, and learning rate are tuned using \textit{Optuna}~\cite{akiba2019optuna}.
In these experiments, we set the threshold policy $\mathcal{T}$ to maximize the F1 score in the classification processes. 
\\
\textbf{Compared Benchmarks}
All compered benchmarks below were trained using the same dataset and evaluated on the same test set.
(1) Isolation Forest (IF), an anomaly detection algorithm that isolates outliers by recursively partitioning data and scoring it based on the number of splits required to isolate an observation~\cite{liu2008isolation}.
(2) One-Class SVM (OCSVM), which learns a decision boundary to separate benign data from outliers, treating all training data as belonging to one class~\cite{li2003improving}.
(3) Autoencoder (AE), which learns to compress and reconstruct data using linear layers, with anomalies detected by measuring reconstruction error, assuming that benign data have lower errors than the anomalous data.

\subsection{Experimental Results}

\subsubsection{Attack}

The experimental results of the \attackname attack are presented in~\Cref{fig:attack_scenarios} and summarized in~\Cref{tab:attacks_sum}. 
\Cref{fig:attack_scenarios} presents the network state for both attack scenarios and the corresponding benign scenarios detailed in~\Cref{subsec:settings}:
(1) SAS with BS5 as the malicious cell (\Cref{fig:attack_scenarios:SAS}), and (2) the MAS with both BS1 and BS5 as malicious cells (\Cref{fig:attack_scenarios:MAS}).
Each cell (BS1-BS6) is presented in a different color, with the y-axis representing the amount of connected UE during each iteration.
\Cref{tab:attacks_sum} summarizes the two scenarios for each cell. 
Each row represents a cell, while the columns represent the different scenarios; for each scenario, presented the average amount of UE served, the percentage difference from the benign scenario, and minimum and maximum UE counts.

When examining the results regarding the SAS, we see a significant increase of 248.5\% in the average amount of UE served by the malicious cell BS5 in the malicious scenario compared to the benign scenario.
Additionally, we observe a higher minimum number of serving UE for BS5 in the malicious scenario, indicating that fewer UE left compared to the benign scenario.
In the MAS, we see the attack's impact across the entire network. 
The malicious cells BS1 and BS5 increased the amount of their served UE by 154.16\% and 248.5\% respectively, while their neighbor cell BS6 suffered a 50.39\% reduction in its average amount of served UE.

\Cref{fig:attack_scenarios} demonstrates the impact of the attack by comparing UE distribution patterns across cells over time. 
In the SAS shown in~\Cref{fig:attack_scenarios:SAS}, where malicious cell BS5 (purple) executes the attack, we observe different behavior between the benign and malicious scenarios. 
While the benign scenario shows BS5's UE count decreasing over time, the malicious scenario shows a significant increase in its UE allocations. 
The MAS shown in~\Cref{fig:attack_scenarios:MAS}, where both BS1 (blue) and BS5 (purple) execute the attack, shows several distinct patterns. 
In the benign scenario, both BS1 and BS5 maintain relatively stable UE counts. 
However, during the attack scenario, both malicious cells demonstrate increases in their UE allocations. 
Notably, this attack significantly impacts neighboring cell BS6 (brown), which experiences a substantial reduction in UE connections compared to its high allocation in the benign scenario.

\begin{figure}[] 
    \centering
    \begin{subfigure}{0.23\textwidth} 
        \centering
        
        \includegraphics[width=\textwidth]
        {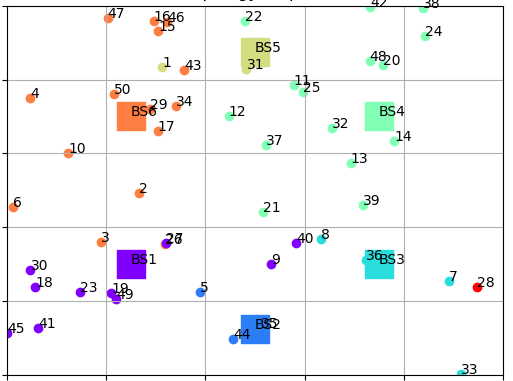}
        \caption{Network topology during benign scenario}
        
        \label{fig:bening}
    \end{subfigure}
    \hfill
    \begin{subfigure}{0.23\textwidth}
        \centering
        \includegraphics[width=\textwidth]{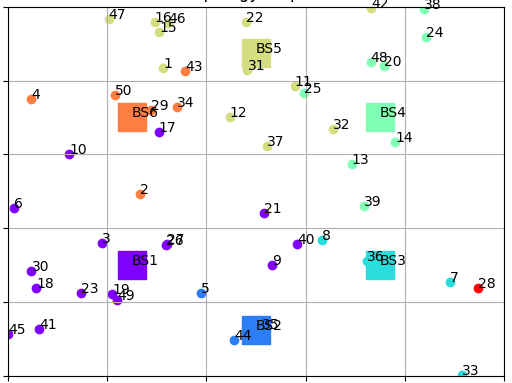} 
        \caption{Network topology during MAS}
        
        \label{fig:MAS}
    \end{subfigure}
    \caption{Network topology during the benign scenario (\Cref{fig:bening}) and MAS (\Cref{fig:MAS}).
    The boxes (BS1 - BS6) are the cells, and the circles (1-50) are the UE IDs.
    The colors represent the UE's association to a cell. 
    Both~\Cref{fig:bening} and ~\Cref{fig:MAS} provide a 
    snapshot of the same simulation iteration.
    }
    \label{fig:combined}
\end{figure}

\Cref{fig:combined} presents the network states for two scenarios: the benign scenario (\Cref{fig:bening}) and the MAS (\Cref{fig:MAS}). 
In both figures, circles represent UE, and boxes represent cells, with each UE's color indicating its serving cell. 
Both \Cref{fig:bening} and \Cref{fig:MAS} provide snapshots for a specific simulation iteration. 
The figures illustrate the impact of the attack, which affects not only the attacker’s cell but also neighboring cells, particularly BS6 (orange), by reducing the amount of UE it serves.

\subsubsection{Detection}

\begin{figure*}[h]
    \centering
    
    \includegraphics[width=1\linewidth]{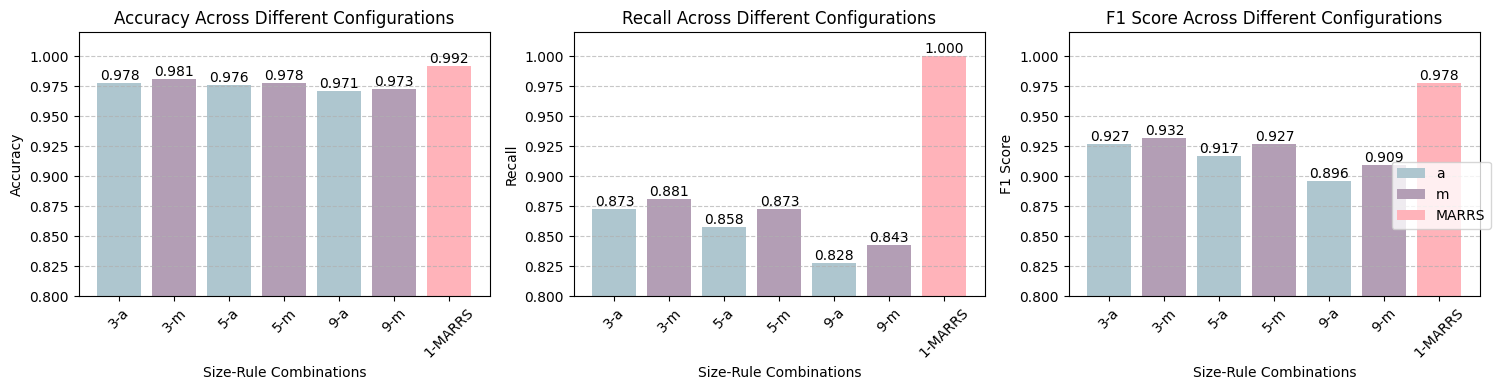}
    \caption{Sequence-based detection approach results.}
    \label{fig:seq}
\end{figure*}

To evaluate \detectname method, we first demonstrate the importance of gathering data over time in the simulation testbed.
The results are presented in~\Cref{tab:evaltime}, where each row presents different subsets of the training set that were used ($x_1, x_2, x_3, x_4$), and the columns present the accuracy, precision, recall, and F1 scores.  
As can be see in the table, the more time the system operates, the more data it collects; accordingly, the classification metrics results improve.
\begin{table}[ht!]
\small 
\centering
\caption{Accuracy of the \detectname method over time.}
    \begin{tabular}{c|cccc}
    \toprule
    \textbf{Training Set} & \textbf{Accuracy} & \textbf{Precision} & \textbf{Recall} & \textbf{F1 Score}\\ 
    \midrule
    $x_1$ &
    0.873 &
    0.793 &
    1 &
    0.884 \\
    \midrule
    $x_1, x_2$ &
    0.984 &
    0.986 &
    0.922 &
    0.949 \\
    \midrule
    $x_1, x_2, x_3$ &
    0.964 &
    0.932 &
    1 &
    0.965 \\
    \midrule
    $x_1, x_2, x_3, x_4$ &
    0.992 &
    0.958 &
    1 &
    0.978 \\
    \bottomrule
    \end{tabular}
    \label{tab:evaltime}
\end{table}
\\
\textbf{Compared Benchmarks}
We compared \detectname's performance to that of other detection methods as detailed in~\Cref{subsec:settings}. 
The results are presented in~\Cref{tab:benchmarks}, where each row represents a method used to detect the \attackname attack, and the columns contain the classification metric values on the test set. 
As can be seen, \detectname outperforms all other methods on the F1 score and accuracy metrics.

\begin{table}[htbp]
\small
\centering
\caption{Performance of the examined detection methods.}
\begin{tabular}{l|l|cccc}
\toprule
\multirow{2}{*}{} & \textbf{Method} & \textbf{Accuracy} & \textbf{Precision} & \textbf{Recall} & \textbf{F1} \\
\midrule
\multirow{3}{*}[1.2pt]{\rotatebox[origin=c]{90}{Benchmarks}} 
& IF & 0.837 & 0.522 & 1 & 0.69 \\
\cmidrule{2-6}
& OCSVM & 0.871 & 0.578 & 0.985 & 0.730 \\
\cmidrule{2-6}
& LAE & 0.873 & 0.793 & 1 & 0.884 \\
\midrule
\multirow{1}{*}{} & {MARRS} & {0.992} & {0.958} & {1} & {0.978} \\
\bottomrule
\end{tabular}
\label{tab:benchmarks}
\end{table}

\noindent\textbf{Ablation Study}
In ablation studies, components of an ML model are systematically removed or altered to assess their impact on the model's overall performance.
The goal is to determine how each part contributes to the overall effectiveness of the model~\cite{molinari2020designing, meyes2019ablation}.
We employed this evaluation process to examine the effectiveness of \detectname's architecture (see~\Cref{fig:AE-LSTM}). 
In the first experiment of the ablation study just the first layer of AEs ($AE^{<1>}$) was used to encode and reconstruct network KPIs and classify them based on their reconstruction loss.
In the second experiment, we trained the AE for each cell using the concatenated average features from the rest of the network along with the cell's own feature set ($AE^{<1+>}$).
This is in contrast to our detection method which also employs two additional steps: latent space feature creation and enriched AE training (described in~\Cref{sec:frame}).

The results are summarized in~\Cref{tab:evalabl}, where each row represents the experiment ($AE^{<1>}$ and $AE^{<1+>}$), and the columns contain the classification metric values.
The results show how incorporating latent space features as contextual information from the entire network improves the detection of malicious cell reports.
\begin{table}[!ht]
\small 
\centering
\caption{Ablation study results.}
    \begin{tabular}{c|cccc}
    \toprule
    \textbf{Layer} & \textbf{Accuracy} & \textbf{Precision} & \textbf{Recall} & \textbf{F1 Score}\\ 
    \midrule
     $AE^{<1>}$ &
    0.917 &
    0.932 &
    1 &
    0.964 \\
    
    \midrule
     $AE^{<1+>}$ &
    0.966 &
    0.934&
    1 &
    0.965 \\
    
    \midrule

    \detectname &
    0.992 &
    0.958 &
    1 &
    0.978 \\
    \bottomrule
    \end{tabular}
    \label{tab:evalabl}
\end{table}

\noindent\textbf{Sequence-Based Detection (S-MARRS)}
\Cref{fig:seq} presents the results of our sequence-based detection approach (see~\Cref{subsec:sbd}), which reduces the false positives that may result from outliers and focuses on malicious behavior by examining a sequence of KPI time windows instead of an individual time window.
The classification rules used in this experiment are as follows:
A (all rule): The sequence is classified as untrusted if all windows within the sequence exceed the threshold $\mathcal{T}$; otherwise, it is classified as trusted.
M (majority rule): The sequence is classified as untrusted if the majority of windows within the sequence exceed the threshold $\mathcal{T}$; otherwise, it is classified as trusted.

Note that on the x-axes in~\Cref{fig:seq}, the number before the letters "A" or "M"  indicates the sequence size.
The evaluation results for accuracy, recall, and F1 score are summarized in~\Cref{fig:seq}.
In terms of precision, all configurations of "A" and "M" obtained a perfect score of 1, however when a sequence-based configuration was not used, which we refer to as the 1-MARRS (see~\Cref{fig:seq} the last column) detection approach, precision of 0.958 was obtained.
The primary goal of adopting a sequence-based detection approach is to minimize the FPR, and the results demonstrate that we have achieved this objective. 
The FPR directly impacts precision and indirectly influences metrics like recall and the F1 score. 
In configurations where precision is a perfect 1, this indicates the absence of false positives, i.e., benign time windows were not misclassified as untrusted. 
On the other hand, the non-sequence detection approach achieved a perfect recall of 1 but at the expense of a higher FPR, which reduced its precision compared to the sequence-based configurations. 
This highlights the tradeoff between high recall and precision when the FPR is not adequately controlled.

\section{Discussion}\label{sec:discussion}

Throughout this paper, we have examined the security vulnerabilities arising from multi-operator deployments.
We have demonstrated the need for robust security solutions and regulatory constraints, compliance, and auditing in this rapidly evolving domain.
According to the introduced threat model presented in~\cref{sec:tm}, we demonstrated how malicious cells can exploit the multi-operator network to manipulate the TS flow and unfairly increase their UE allocations (the \attackname attack). 
The risks that arise from applying the \attackname attack, not only undermine the integrity of the network but also degrades the QoE for UEs, presenting new  security challenges for the O-RAN architecture.

However, the presented threat model (\cref{sec:tm}) is not limited to scenarios that only involve malicious operators as threat actors. 
A similar threat can emerge even in single-operator networks if a cell's supply chain is compromised. 
An attacker who gains control of a cell via a supply chain breach could execute the \attackname attack, resulting in disruptions similar to those detailed in~\Cref{sec:eval} such as QoE reduction, which could harm the operator's reputation.
Deploying MARSS on the near-RT RIC offers a solution to these challenges as well, as it treats all telemetry as untrusted, enabling the detection and mitigation of such threats.

\section{Conclusion and Future Work}\label{sec:conc}

This paper addresses O-RAN vulnerabilities in multi-operator environments. 
We introduce the \attackname to demonstrate how a malicious operator can exploit these vulnerabilities by executing an evasion attack that misleads O-RAN traffic steering and disrupts network load distribution. 
To counter such threats and ensure the continuation of O-RAN's legitimate operation, we developed the \detectname framework, designed to detect compromised telemetry in real time.
Our evaluation reveals \detectname achieves high precision, recall, and F1 scores in detecting malicious cell behavior.

Future work can focus on enhancing the \detectname framework by developing an innovative approach for automated policy selection using deep reinforcement learning (DRL). 
Currently, the policies (\Cref{subsec:class}) in \detectname must be manually managed by the operators, a process which may be prone to human error.
However, leveraging the RIC architecture, it is possible to train a DRL-based policy-making model within the service management and orchestration (SMO) layer. 
This model would be able to be deployed as rApp on the non-RT RIC and integrated with the \detectname xApp hosted on the near-RT RIC via the A1 interface. 
This automated approach could significantly improve the proposed \detectname's adaptability and efficiency while reducing human involvement.

\bibliographystyle{plain}
\bibliography{reference}

\begin{thebibliography}{10}

\bibitem{3gpp2013ran}
{3rd Generation Partnership Project}.
\newblock {TR 22.852: Study on RAN Sharing Enhancements (Release 12 \& 13)}.
\newblock Technical report, Technical Specification Group Radio Access Networks, 2013.

\bibitem{ahmad2018overview}
Ijaz Ahmad, Tanesh Kumar, Madhusanka Liyanage, Jude Okwuibe, Mika Ylianttila, and Andrei Gurtov.
\newblock Overview of 5g security challenges and solutions.
\newblock {\em IEEE Communications Standards Magazine}, 2(1):36--43, 2018.

\bibitem{akiba2019optuna}
Takuya Akiba, Shotaro Sano, Toshihiko Yanase, Takeru Ohta, and Masanori Koyama.
\newblock Optuna: A next-generation hyperparameter optimization framework.
\newblock In {\em Proceedings of the 25th ACM SIGKDD international conference on knowledge discovery \& data mining}, pages 2623--2631, 2019.

\bibitem{alimohammadikpi}
Hamed Alimohammadi, Sotiris Chatzimiltis, Samara Mayhoub, Mohammad Shojafar, Seyed~Ahmad Soleymani, Ayhan Akbas, and Chuan~Heng Foh.
\newblock Kpi poisoning: An attack in open ran near real-time control loop.

\bibitem{WG11}
O-RAN ALLIANCE.
\newblock {WG11: O-RAN Work Group 11 (Security Work Group) Security Requirements and Controls Specifications}, 2024.

\bibitem{azariah2024survey}
Wilfrid Azariah, Fransiscus~Asisi Bimo, Chih-Wei Lin, Ray-Guang Cheng, Navid Nikaein, and Rittwik Jana.
\newblock A survey on open radio access networks: Challenges, research directions, and open source approaches.
\newblock {\em Sensors}, 24(3):1038, 2024.

\bibitem{baguer2024attacking}
Pau Baguer, Girma~M Yilma, Esteban Municio, Gines Garcia-Aviles, Andres Garcia-Saavedra, Marco Liebsch, and Xavier Costa-P{\'e}rez.
\newblock Attacking o-ran interfaces: Threat modeling, analysis and practical experimentation.
\newblock {\em IEEE Open Journal of the Communications Society}, 2024.

\bibitem{balakrishnan2024enhancing}
Ravikumar Balakrishnan, Marius Arvinte, Nageen Himayat, Hosein Nikopour, and Hassnaa Moustafa.
\newblock Enhancing o-ran security: Evasion attacks and robust defenses for graph reinforcement learning-based connection management.
\newblock {\em arXiv preprint arXiv:2405.03891}, 2024.

\bibitem{balasubramanian2021ric}
Bharath Balasubramanian, E~Scott Daniels, Matti Hiltunen, Rittwik Jana, Kaustubh Joshi, Rajarajan Sivaraj, Tuyen~X Tran, and Chengwei Wang.
\newblock Ric: A ran intelligent controller platform for ai-enabled cellular networks.
\newblock {\em IEEE Internet Computing}, 25(2):7--17, 2021.

\bibitem{biggio2013evasion}
Battista Biggio, Igino Corona, Davide Maiorca, Blaine Nelson, Nedim {\v{S}}rndi{\'c}, Pavel Laskov, Giorgio Giacinto, and Fabio Roli.
\newblock Evasion attacks against machine learning at test time.
\newblock In {\em Machine Learning and Knowledge Discovery in Databases: European Conference, ECML PKDD 2013, Prague, Czech Republic, September 23-27, 2013, Proceedings, Part III 13}, pages 387--402. Springer, 2013.

\bibitem{bimo2022osc}
Fransiscus~Asisi Bimo, Ferlinda Feliana, Shu-Hua Liao, Chih-Wei Lin, David~F Kinsey, James Li, Rittwik Jana, Richard Wright, and Ray-Guang Cheng.
\newblock Osc community lab: The integration test bed for o-ran software community.
\newblock In {\em 2022 IEEE Future Networks World Forum (FNWF)}, pages 513--518. IEEE, 2022.

\bibitem{bitton2023evaluating}
Ron Bitton, Nadav Maman, Inderjeet Singh, Satoru Momiyama, Yuval Elovici, and Asaf Shabtai.
\newblock Evaluating the cybersecurity risk of real-world, machine learning production systems.
\newblock {\em ACM Computing Surveys}, 55(9):1--36, 2023.

\bibitem{bonati2020open}
Leonardo Bonati, Michele Polese, Salvatore D’Oro, Stefano Basagni, and Tommaso Melodia.
\newblock Open, programmable, and virtualized 5g networks: State-of-the-art and the road ahead.
\newblock {\em Computer Networks}, 182:107516, 2020.

\bibitem{branco2024evaluation}
Paulo~Ricardo Branco~da Silva, Jo{\~a}o~Paulo Henriques Sales~de Lima, Erika Costa~Alves, William~Sanchez Farfan, Victor~Aguiar Coutinho, Thomas William do~Prado Paiva, Daniel~Lazkani Feferman, and Francisco Hugo~Costa Neto.
\newblock Evaluation of the latency of machine learning random access ddos detection in open ran.
\newblock In {\em Proceedings of the 30th Annual International Conference on Mobile Computing and Networking}, pages 2306--2311, 2024.

\bibitem{chen2023flexslice}
Chieh-Chun Chen, Chia-Yu Chang, and Navid Nikaein.
\newblock Flexslice: Flexible and real-time programmable ran slicing framework.
\newblock In {\em GLOBECOM 2023-2023 IEEE Global Communications Conference}, pages 3807--3812. IEEE, 2023.

\bibitem{chen2020hopskipjumpattack}
Jianbo Chen, Michael~I Jordan, and Martin~J Wainwright.
\newblock Hopskipjumpattack: A query-efficient decision-based attack.
\newblock In {\em 2020 ieee symposium on security and privacy (sp)}, pages 1277--1294. IEEE, 2020.

\bibitem{de2022satellite}
Emanuele De~Santis, Alessandro Giuseppi, Antonio Pietrabissa, Michael Capponi, and Francesco Delli~Priscoli.
\newblock Satellite integration into 5g: deep reinforcement learning for network selection.
\newblock {\em Machine Intelligence Research}, 19(2):127--137, 2022.

\bibitem{demestichas20135g}
Panagiotis Demestichas, Andreas Georgakopoulos, Dimitrios Karvounas, Kostas Tsagkaris, Vera Stavroulaki, Jianmin Lu, Chunshan Xiong, and Jing Yao.
\newblock 5g on the horizon: Key challenges for the radio-access network.
\newblock {\em IEEE vehicular technology magazine}, 8(3):47--53, 2013.

\bibitem{dryjanski2021toward}
Marcin Dryja{\'n}ski, {\L}ukasz Ku{\l}acz, and Adrian Kliks.
\newblock Toward modular and flexible open ran implementations in 6g networks: Traffic steering use case and o-ran xapps.
\newblock {\em Sensors}, 21(24):8173, 2021.

\bibitem{keith2024oran}
Keith Dyer.
\newblock Freshwave says four way operator sharing on same indoor small cells a world first, 2024.

\bibitem{esa2024oran}
{ESA}.
\newblock {Spacetime and O-RAN Interfaces 5G/6G NTN}, 2024.
\newblock https://connectivity.esa.int/projects/spacetime-and-oran-interfaces-5g6g-ntns.

\bibitem{farhat2015access}
Soha Farhat, Zahraa Chahine, Abed~Ellatif Samhat, Samer Lahoud, and Bernard Cousin.
\newblock Access selection and joint pricing in multi-operator wireless networks: A stackelberg game.
\newblock In {\em 2015 Fifth International Conference on Digital Information and Communication Technology and its Applications (DICTAP)}, pages 38--43. IEEE, 2015.

\bibitem{farhat2014best}
Soha Farhat, Abed~Ellatif Samhat, Samer Lahoud, and Bernard Cousin.
\newblock Best operator policy in a heterogeneous wireless network.
\newblock In {\em The Third International Conference on e-Technologies and Networks for Development (ICeND2014)}, pages 53--57. IEEE, 2014.

\bibitem{farhat2017radio}
Soha Farhat, Abed~Ellatif Samhat, Samer Lahoud, and Bernard Cousin.
\newblock Radio access network sharing in 5g: strategies and benefits.
\newblock {\em Wireless Personal Communications}, 96:2715--2740, 2017.

\bibitem{goodfellow2014explaining}
Ian~J Goodfellow.
\newblock Explaining and harnessing adversarial examples.
\newblock {\em arXiv preprint arXiv:1412.6572}, 2014.

\bibitem{groen2024implementing}
Joshua Groen, Salvatore D’Oro, Utku Demir, Leonardo Bonati, Michele Polese, Tommaso Melodia, and Kaushik Chowdhury.
\newblock Implementing and evaluating security in o-ran: Interfaces, intelligence, and platforms.
\newblock {\em IEEE Network}, 2024.

\bibitem{vodo2019oran}
Vodafone Group.
\newblock Vodafone announces expanded network sharing agreement with orange in spain, 2019.

\bibitem{gsma2018}
{GSMA}.
\newblock Mobile infrastructure sharing.

\bibitem{habler2022adversarial}
Edan Habler, Ron Bitton, Dan Avraham, Dudu Mimran, Eitan Klevansky, Oleg Brodt, Heiko Lehmann, Yuval Elovici, and Asaf Shabtai.
\newblock Adversarial machine learning threat analysis and remediation in open radio access network (o-ran).
\newblock {\em arXiv preprint arXiv:2201.06093}, 2022.

\bibitem{hasabelnaby2024centralized}
Mahmoud~A Hasabelnaby, Mohanad Obeed, Mohammed Saif, Anas Chaaban, and MJ~Hossain.
\newblock From centralized ran to open ran: A survey on the evolution of distributed antenna systems.
\newblock {\em arXiv preprint arXiv:2411.12166}, 2024.

\bibitem{hung2024security}
Cheng-Feng Hung, You-Run Chen, CHI-Heng Tseng, and Shin-Ming Cheng.
\newblock Security threats to xapps access control and e2 interface in o-ran.
\newblock {\em IEEE Open Journal of the Communications Society}, 2024.

\bibitem{eugina2024ric}
Eugina Jordan.
\newblock Ric: The next phase of open ran, 2024.

\bibitem{kee2021oran}
Shin~Yuan Kee.
\newblock How does open ran add value in multi-operator sharing?, 2021.

\bibitem{lee2024lstm}
Younjeong Lee, Chanho Park, Namji Kim, Jisu Ahn, and Jongpil Jeong.
\newblock Lstm-autoencoder based anomaly detection using vibration data of wind turbines.
\newblock {\em Sensors}, 24(9):2833, 2024.

\bibitem{li2003improving}
Kun-Lun Li, Hou-Kuan Huang, Sheng-Feng Tian, and Wei Xu.
\newblock Improving one-class svm for anomaly detection.
\newblock In {\em Proceedings of the 2003 international conference on machine learning and cybernetics (IEEE Cat. No. 03EX693)}, volume~5, pages 3077--3081. IEEE, 2003.

\bibitem{liu2008isolation}
Fei~Tony Liu, Kai~Ming Ting, and Zhi-Hua Zhou.
\newblock Isolation forest.
\newblock In {\em 2008 eighth ieee international conference on data mining}, pages 413--422. IEEE, 2008.

\bibitem{liu2024democratizing}
Lixin Liu, Yuanjie Li, Hewu Li, Jiabo Yang, Wei Liu, Jingyi Lan, Yufeng Wang, Jiarui Li, Jianping Wu, Qian Wu, et~al.
\newblock Democratizing $\{$Direct-to-Cell$\}$ low earth orbit satellite networks.
\newblock In {\em 21st USENIX Symposium on Networked Systems Design and Implementation (NSDI 24)}, pages 791--808, 2024.

\bibitem{liyanage2023open}
Madhusanka Liyanage, An~Braeken, Shahriar Shahabuddin, and Pasika Ranaweera.
\newblock Open ran security: Challenges and opportunities.
\newblock {\em Journal of Network and Computer Applications}, 214:103621, 2023.

\bibitem{dtdf2021oran}
Johannes Maisack.
\newblock Deutsche telekom and telefónica share network infrastructure to enhance network coverage, 2021.

\bibitem{marinova2024intelligent}
Simona Marinova and Alberto Leon-Garcia.
\newblock Intelligent o-ran beyond 5g: Architecture, use cases, challenges, and opportunities.
\newblock {\em IEEE Access}, 12:27088--27114, 2024.

\bibitem{markendahl2013shared}
Jan Markendahl and Amirhossein Ghanbari.
\newblock Shared smallcell networks multi-operator or third party solutions-or both?
\newblock In {\em 2013 11th International symposium and workshops on modeling and optimization in mobile, Ad Hoc and wireless networks (WiOpt)}, pages 41--48. IEEE, 2013.

\bibitem{markendahl2013network}
Jan Markendahl, Amirhossein Ghanbari, and Bengt~G M{\"o}lleryd.
\newblock Network cooperation between mobile operators-why and how competitors cooperate?
\newblock In {\em IMP conf, Atlanta}, 2013.

\bibitem{meyes2019ablation}
Richard Meyes, Melanie Lu, Constantin~Waubert de~Puiseau, and Tobias Meisen.
\newblock Ablation studies in artificial neural networks.
\newblock {\em arXiv preprint arXiv:1901.08644}, 2019.

\bibitem{mimran2022evaluating}
Dudu Mimran, Ron Bitton, Yehonatan Kfir, Eitan Klevansky, Oleg Brodt, Heiko Lehmann, Yuval Elovici, and Asaf Shabtai.
\newblock Evaluating the security of open radio access networks.
\newblock {\em arXiv preprint arXiv:2201.06080}, 2022.

\bibitem{molinari2020designing}
Alessio Molinari.
\newblock Designing a performant ablation study framework for pytorch, 2020.

\bibitem{nec2013ran}
{NEC Corporation}.
\newblock Ran sharing: Nec’s approach towards active radio access network sharing.
\newblock Techreport, NEC Corporation, 2013.

\bibitem{nicolae2018adversarial}
Maria-Irina Nicolae, Mathieu Sinn, Minh~Ngoc Tran, Beat Buesser, Ambrish Rawat, Martin Wistuba, Valentina Zantedeschi, Nathalie Baracaldo, Bryant Chen, Heiko Ludwig, et~al.
\newblock Adversarial robustness toolbox v1. 0.0.
\newblock {\em arXiv preprint arXiv:1807.01069}, 2018.

\bibitem{niknam2022intelligent}
Solmaz Niknam, Abhishek Roy, Harpreet~S Dhillon, Sukhdeep Singh, Rahul Banerji, Jeffery~H Reed, Navrati Saxena, and Seungil Yoon.
\newblock Intelligent o-ran for beyond 5g and 6g wireless networks.
\newblock In {\em 2022 IEEE Globecom Workshops (GC Wkshps)}, pages 215--220. IEEE, 2022.

\bibitem{WG2}
{O-RAN ALLIANCE}.
\newblock {WG2: Non-real-time RAN Intelligent Controller and A1 Interface Workgroup}, 2022.

\bibitem{WG3}
{O-RAN ALLIANCE}.
\newblock {WG3: Near-real-time RIC and E2 Interface Workgroup}, 2022.

\bibitem{opadere2019energy}
Johnson Opadere, Qiang Liu, Tao Han, and Nirwan Ansari.
\newblock Energy-efficient virtual radio access networks for multi-operators cooperative cellular networks.
\newblock {\em IEEE Transactions on Green Communications and Networking}, 3(3):603--614, 2019.

\bibitem{orhan2021connection}
Oner Orhan, Vasuki~Narasimha Swamy, Thomas Tetzlaff, Marcel Nassar, Hosein Nikopour, and Shilpa Talwar.
\newblock Connection management xapp for o-ran ric: A graph neural network and reinforcement learning approach.
\newblock In {\em 2021 20th IEEE International Conference on Machine Learning and Applications (ICMLA)}, pages 936--941. IEEE, 2021.

\bibitem{park2024investigation}
Heejae Park, Tri-Hai Nguyen, and Laihyuk Park.
\newblock An investigation on open-ran specifications: Use cases, security threats, requirements, discussions.
\newblock {\em CMES-Computer Modeling in Engineering \& Sciences}, 141(1), 2024.

\bibitem{parvez2018survey}
Imtiaz Parvez, Ali Rahmati, Ismail Guvenc, Arif~I Sarwat, and Huaiyu Dai.
\newblock A survey on low latency towards 5g: Ran, core network and caching solutions.
\newblock {\em IEEE Communications Surveys \& Tutorials}, 20(4):3098--3130, 2018.

\bibitem{paszke2017automatic}
Adam Paszke, Sam Gross, Soumith Chintala, Gregory Chanan, Edward Yang, Zachary DeVito, Zeming Lin, Alban Desmaison, Luca Antiga, and Adam Lerer.
\newblock Automatic differentiation in pytorch.
\newblock 2017.

\bibitem{polese2023understanding}
Michele Polese, Leonardo Bonati, Salvatore D’oro, Stefano Basagni, and Tommaso Melodia.
\newblock Understanding o-ran: Architecture, interfaces, algorithms, security, and research challenges.
\newblock {\em IEEE Communications Surveys \& Tutorials}, 25(2):1376--1411, 2023.

\bibitem{priscoli2020traffic}
Francesco~Delli Priscoli, Alessandro Giuseppi, Francesco Liberati, and Antonio Pietrabissa.
\newblock Traffic steering and network selection in 5g networks based on reinforcement learning.
\newblock In {\em 2020 European Control Conference (ECC)}, pages 595--601. IEEE, 2020.

\bibitem{sagheer2019unsupervised}
Alaa Sagheer and Mostafa Kotb.
\newblock Unsupervised pre-training of a deep lstm-based stacked autoencoder for multivariate time series forecasting problems.
\newblock {\em Scientific reports}, 9(1):19038, 2019.

\bibitem{nisar2023}
Nisar Sanadi.
\newblock How will ric leverage ai/ml to improve user experience?, December 2023.

\bibitem{sapavath2023experimental}
Naveen~Naik Sapavath, Brian Kim, Kaushik Chowdhury, and Vijay~K Shah.
\newblock Experimental study of adversarial attacks on ml-based xapps in o-ran.
\newblock {\em arXiv preprint arXiv:2309.03844}, 2023.

\bibitem{shen2022security}
Chih-Ting Shen, Yu-Yi Xiao, Yi-Wei Ma, Jiann-Liang Chen, Cheng-Mou Chiang, Shiang-Jiun Chen, and Yu-Chuan Pan.
\newblock Security threat analysis and treatment strategy for oran.
\newblock In {\em 2022 24th International Conference on Advanced Communication Technology (ICACT)}, pages 417--422. IEEE, 2022.

\bibitem{soleymaniddos}
Seyed~Ahmad Soleymani, Mohsen Eslamnejad, Hamed Alimohammadi, Ayhan Akbas, Chuan~Heng Foh, and Mohammad Shojafar.
\newblock Ddos detection and mitigation using d/xapp in o-ran.

\bibitem{soltani2024intelligent}
Sanaz Soltani, Mohammad Shojafar, Ali Amanlou, and Rahim Tafazolli.
\newblock Intelligent control in 6g open ran: Security risk or opportunity?
\newblock {\em arXiv preprint arXiv:2405.08577}, 2024.

\bibitem{otto2021oran}
Otto T.
\newblock Neutral host: how open ran and neutral host paves the way for 5g, 2021.

\bibitem{tabassi2019taxonomy}
Elham Tabassi, Kevin~J Burns, Michael Hadjimichael, Andres~D Molina-Markham, and Julian~T Sexton.
\newblock A taxonomy and terminology of adversarial machine learning.
\newblock {\em NIST IR}, 2019:1--29, 2019.

\bibitem{tayyab2019survey}
Muhammad Tayyab, Xavier Gelabert, and Riku J{\"a}ntti.
\newblock A survey on handover management: From lte to nr.
\newblock {\em IEEE Access}, 7:118907--118930, 2019.

\bibitem{towergenius2018}
{Tower Genius LLC}.
\newblock Cell tower co-location.
\newblock \url{https://www.cell-phone-towers.com/Cell-Tower-Colocation.html}, 2018.
\newblock Accessed: Jul. 30, 2018.

\bibitem{tsourdinis2024ai}
Theodoros Tsourdinis, Nikos Makris, Thanasis Korakis, and Serge Fdida.
\newblock Ai-driven network intrusion detection and resource allocation in real-world o-ran 5g networks.
\newblock In {\em Proceedings of the 30th Annual International Conference on Mobile Computing and Networking}, pages 1842--1849, 2024.

\bibitem{vodafone2024}
Vodafone.
\newblock Technology and innovation open ran, 2024.

\bibitem{wei2023lstm}
Yuanyuan Wei, Julian Jang-Jaccard, Wen Xu, Fariza Sabrina, Seyit Camtepe, and Mikael Boulic.
\newblock Lstm-autoencoder-based anomaly detection for indoor air quality time-series data.
\newblock {\em IEEE Sensors Journal}, 23(4):3787--3800, 2023.

\bibitem{wooden2024oran}
Andrew Wooden.
\newblock Freshwave pumps out 4g from all four uk operators in one unit, 2024.

\bibitem{xavier2023machine}
Bruno~Missi Xavier, Merim Dzaferagic, Diarmuid Collins, Giovanni Comarela, Magnos Martinello, and Marco Ruffini.
\newblock Machine learning-based early attack detection using open ran intelligent controller.
\newblock {\em arXiv preprint arXiv:2302.01864}, 2023.

\bibitem{xavier2024cross}
Bruno~Missi Xavier, Merim Dzaferagic, Irene Vil{\`a}, Magnos Martinello, and Marco Ruffini.
\newblock Cross-domain ai for early attack detection and defense against malicious flows in o-ran.
\newblock {\em arXiv preprint arXiv:2401.09204}, 2024.

\bibitem{zhang2020robust}
Huan Zhang, Hongge Chen, Chaowei Xiao, Bo~Li, Mingyan Liu, Duane Boning, and Cho-Jui Hsieh.
\newblock Robust deep reinforcement learning against adversarial perturbations on state observations.
\newblock {\em Advances in Neural Information Processing Systems}, 33:21024--21037, 2020.

\bibitem{zhu2021integrated}
Xiangming Zhu and Chunxiao Jiang.
\newblock Integrated satellite-terrestrial networks toward 6g: Architectures, applications, and challenges.
\newblock {\em IEEE Internet of Things Journal}, 9(1):437--461, 2021.

\end{thebibliography}

\end{document}